\documentclass[english,showpacs,floatfix,12pt]{revtex4}
\usepackage[dvips]{graphicx}
\usepackage{longtable}
\usepackage{amsmath,amssymb}
\usepackage{dcolumn}
\usepackage{latexsym}
\usepackage{babel}
\usepackage[latin1]{inputenc}
\usepackage{color}
\usepackage[colorlinks]{hyperref}
\def\al{\alpha}

\def\kp{\kappa}

\def\pa{\partial}
\def\vf{\varphi}

\def\Om{\Omega}

\def\ga{\gamma}
\def\be{\beta}
\def\dl{\delta}

\def\th{\theta}

\def\E{{\cal E}}

\def\diag{\mbox {diag}}

\def\sgn{\mbox {sgn}}
\def\ln{\mbox {ln}}

\def\l{\left}
\def\r{\right}

\begin{document}
\title{\Large The Exact Solutions to the Gravitational Contraction
\\ in Comoving Coordinate System}
\author{Ying-Qiu Gu}
\email{yqgu@fudan.edu.cn} \affiliation{School of Mathematical
Science, Fudan University, Shanghai 200433, China} 
\date{15th March 2009}

\begin{abstract}
The gravitational collapse of a star is a warmly discussed but
still puzzling problem, which not only involves the dynamics of
the gases, but also the subtle coordinate transformation. In this
letter, we give some more detailed investigation on this problem,
and reach the results: (I). The comoving coordinate system for the
stellar system is only compatible with the zero-pressure free
falling particles. (II). For the free falling dust, there are
three kind of solutions respectively corresponding to the
oscillating, the critical and the open trajectories. The solution
of Oppenheimer and Snyder is the critical case. (III). All
solutions are exactly derived. There is a new kind singularity in
the solution, but its origin is unclear. \vskip3mm
\noindent{Keywords:} {\sl gravitational collapse, comving
coordinate system, exact solution, singularity}

\end{abstract}
\maketitle

The gravitational collapse of a star is a warmly discussed but
still puzzling problem, which not only involves the dynamics of
the gases, but also the subtle coordinate transformation. The most
enlighten model is the free falling dust model analyzed by J.
Oppenheimer and H. Snyder\cite{Oppen}. In the paper, they said:
``To investigate this question we will solve the field equations
with the limiting form of the energy-momentum tensor in which the
pressure is zero. When the pressure vanishes there are no static
solutions to the field equations except when all components of
$T^\mu_\nu$ vanishes''. ``we are supposing that the relationships
between the $T^\mu_\nu$ do not admit any stationary solutions, and
therefore exclude this possibility.'' So they actually assumed the
singularity exists, and ignored the dominant gravitational
potential itself is a more powerful source of pressure than the
electromagnetic interaction\cite{gu1,gu6}. The gravity is
conservative, so the free falling particles will move outward when
they have passed across the stellar center. In addition, the
global validity of the comoving coordinate system was employed
without a careful check. Actually, the global simultaneity is not
a general valid concept\cite{gu8,gu9}.

In this letter, we give some more detailed investigation on this
problem. From detailed calculation, we find that:
\begin{enumerate}

\item The comoving coordinate system for the stellar system is
only compatible with the zero-pressure free falling particles, so
it may be not a globally valid coordinate system for the case.

\item Similarly to the cosmological model, for the free falling
dust we also have three kind of solutions, namely, the
oscillating, the critical and the open trajectories. The solution
of Oppenheimer and Snyder corresponds to the critical one.

\item All exact solutions to the free falling dust can be
manifestly derived, but the origin of a new kind singularity is
not clarified.

\item The relative equation between the normal spherical
coordinate system and the comoving one is simplified, but it is a
nonlinear hydrodynamical equation, so the global solution might
not exist in the general cases.

\end{enumerate}

The line element in the comoving coordinate system is given by
\begin{eqnarray}
ds^2= d t^2-e^{\bar w}d r^2-e^w(d\th^2+\sin^2\th d\vf^2).
\label{1.1}
\end{eqnarray}
In this case, we have the 4-dimensional speed of the fluid $ U^\mu
=(1,0,0,0)$, and then the energy-momentum tensor becomes
\begin{eqnarray}
T_{\mu\nu} =\diag(\rho,-Pe^{\bar w},-Pe^w,-Pe^w\sin^2\th).
\label{1.2}
\end{eqnarray}
In such coordinate system, we find all exact solutions can be
solved by a bootstrapping method. That is, we firstly express all
variables as functions of $w$ and its derivatives, and then solve
a simplified equation for $w$.

By the continuity equation $U_\mu T^{\mu\nu}_{~;\nu}=0$,
manifestly $ U^\mu \pa_\mu \rho +(\rho+P)U^\mu_{~;\mu}=0$, we get
\begin{eqnarray}
\frac 1 2 \pa_t \bar w +\pa_t w + \frac {\pa_t \rho} {\rho+P} =
0.\label{1.4}
\end{eqnarray}
For the Barotropic fluid, defining
\begin{eqnarray}
B(\rho)=\int \frac {d\rho}{\rho+P(\rho)},\label{1.5}
\end{eqnarray}
then we can solve $\bar w$ from (\ref{1.4})
\begin{eqnarray}
\bar w = -2(w+B)+\al(r),\label{1.6}
\end{eqnarray}
where $\al$ is a function to be determined. Substituting
(\ref{1.6}) into the Einstein equation $G_{tr}+\kp
T_{tr}=0,(\kp=8\pi G)$, namely\cite{Oppen,wein,Tolm}
\begin{eqnarray}
\pa_{tr}^2 w+\frac 1 2 (\pa_t w\pa_r w-\pa_t \bar w \pa_r w) =
0,\label{1.7}
\end{eqnarray}
we get
\begin{eqnarray}
B = -\frac3 2 w-\ln(\pa_r w)+\be(r), \label{1.8}
\end{eqnarray}
where $\be(r)$ is also a function to be determined. Substituting
$B$ into (\ref{1.6}), we get $\bar w$ expressed by $w$
\begin{eqnarray}
\bar w = w+2\ln(\pa_r w)-\ln[4(1+\ga)],\label{1.9}
\end{eqnarray}
in which $\ga(r)=\frac 1 4 e^{2\be-\al}-1$ is introduced for
convenience. The effect of $\ga$ is similar to that of the
dimensionless energy density $\Om$ in cosmology as shown below,
which determines the different evolving fate of the star.

Substituting the above results into $G_{tt}+\kp T_{tt}=0$, we get
$\rho$ expressed by $w$
\begin{eqnarray}
\rho =- \frac 1{\kp\pa_r w}\l[(\ga\pa_r w +2\pa_r \ga)e^{-w}
+\pa_t w (3/4\pa_t w \pa_r w +\pa^2_{tr}w)\r]. \label{1.10}
\end{eqnarray}
Taking $\pa_t \rho$ as variable independent of $\rho$ and
substituting (\ref{1.10}) into $G_{rr}+\kp T_{rr}=0$, we get
$\pa_t \rho$ expressed by $w$
\begin{eqnarray}
\pa_t \rho =\frac 1 {2\kp(\pa_r w)^2} (2\pa^2_{tr} w + 3\pa_t w
\pa_r w ) (\pa^2_t w\pa_r w-\pa_t w \pa^2_{tr} w +2\pa_r \ga
 e^{-w}). \label{1.11}
\end{eqnarray}
Substituting (\ref{1.10}) into (\ref{1.11}), we get the equation
of consistence
\begin{eqnarray}
\pa^3_{t^2 r} w = -\frac 3 2 \pa_t w \pa_{tr} w + \pa_r(\ga
e^{-w}). \label{1.12}
\end{eqnarray}
Integrating (\ref{1.12}) with respect to $r$, we get
\begin{eqnarray}
\pa^2_{t} w = -\frac 3 4 (\pa_t w)^2 +\ga e^{-w}+\dl(t),
\label{1.13}
\end{eqnarray}
where $\dl(t)$ is a function to be determined. In Oppenheimer and
Snyder's work, they assume $\ga=\dl=0$. As calculated below, their
solution corresponds to the critical case.

Substituting (\ref{1.12}) and (\ref{1.13}) into $G_{\th\th}+\kp
T_{\th\th}=0$, we see the physical meaning of $\dl$
\begin{eqnarray}
\dl(t)=-\kp P. \label{1.14}
\end{eqnarray}
Relation (\ref{1.14}) has an important implication, that is, the
metric (\ref{1.1}) is unsuitable to describe the realistic process
of the star contraction, because in this case we definitely have
$\pa_rP\ne 0$, which contradicts (\ref{1.14}). The comoving
coordinate chart is usually a local coordinate chart in heavily
curved space-time, and only in cosmology (\ref{1.1}) is globally
valid, because in this case $P=P(t)$. This shows the limitation of
the comoving metric (\ref{1.1}).

By (\ref{1.14}), we find the only consistent matter in the
space-time describe by (\ref{1.1}) is the fluid with zero
pressure. Of course, this model is extremely idealized, because
the particles passing across the center will continue to move
outward, and the collision among particles leads to high pressure
and temperature. Such high pressure will balance the
contraction\cite{gu1,gu6}.

Now we solve (\ref{1.13}) with $\dl=-\kp P=0$. Setting $w = \frac
4 3 \ln(u)$, by (\ref{1.13}) we get
\begin{eqnarray}
\pa_t^2 u = \frac {3\ga} {4} u^{-\frac1 3},\qquad (\pa_t u)^2 =
\frac {9}{4}\l(\E+\ga u^{\frac 2 3}\r), \label{1.15}
\end{eqnarray}
where ${\E}(r)$ is a function to be determined. Again setting
$u=v^{\frac 3 2}$, i.e. making transformation
\begin{eqnarray}
w = \ln(v^2),\qquad v=e^{\frac w 2}, \end{eqnarray} we get
\begin{eqnarray}
(\pa_t v)^2 = {\frac{\E+\ga v}{v}}. \label{1.16}
\end{eqnarray}

Substituting (\ref{1.16}) into (\ref{1.10}), we learn the physical
meaning of ${\E}$
\begin{eqnarray}
\pa_r{\E}=\kp  \rho v^2 \pa_r v,\qquad \E(r)={\E}_0+\int_0^r\kp
\rho v^2 \pa_r v dr, \label{1.16*}
\end{eqnarray}
where ${\E}_0 $ is a constant. $\E(r)$ corresponds to the comoving
mass in the ball of radius $r$. Therefore, the quantity $\int_0^r
\rho v^2 \pa_r v dr$ is conserved. By their physical meanings, we
certainly have $\pa_r v>0$ and $\pa_r {\E}\ge 0$. From
(\ref{1.16*}), we find $v$ can be regarded as new spatial
coordinate to replace $r$. Substituting $\pa_r v$ into
(\ref{1.9}), we get
\begin{eqnarray}
e^{\frac {\bar w}{2}} =\frac {\pa_r\E}{\kp\rho
v^2\sqrt{1+\ga}},\qquad \rho=\frac {\pa_r\E e^{\frac {-\bar
w}{2}}}{\kp v^2 \sqrt{1+\ga}}.\label{1.19}\end{eqnarray}

Among the interim parameters $(\al,\be,\ga,\dl,{\E})$, only
$(\ga,{\E})$ are independent functions. Their relations are given
by
\begin{eqnarray}
\be=\ln\frac {2\pa_r{\E}}{\kp},\quad (B=\ln\rho),\quad
\al=2\ln\frac{\pa_r{\E}}{\kp\sqrt{1+\ga}},\quad \dl=\kp P,
\end{eqnarray}
where $\be$ is derived by substituting (\ref{1.16}) into
(\ref{1.8}), and $\al$ is derived by the definition of $\ga$.

If $\ga=0$, we get the solution derived in \cite{Oppen} as follows
\begin{eqnarray}
v = \frac {1}{2}\sqrt[3]{18{\E}|t-t_0(r)|^2}. \label{1.16s}
\end{eqnarray}
The curvature at $t=t_0(r)$ is singular. Substituting
(\ref{1.16s}) into (\ref{1.9}), we get
\begin{eqnarray}
e^{\frac {\bar w} 2}=\frac{|\pa_r\E(t-t_0)-2\E\pa_r
t_0|}{\sqrt[3]{{12\E^2}|t-t_0|}} . \label{1.20}
\end{eqnarray}
Substituting (\ref{1.16s}) and (\ref{1.20}) into (\ref{1.19}), we
find
\begin{eqnarray}
\rho = \frac {4 \pa_r \E e^{-\frac {\bar w}2}} {3\kp
\sqrt[3]{12\E^2(t-t_0)^4}}=\frac {4{\pa_r{\E}}}{3\kp
|[{\pa_r{\E}}(t-t_0)-2{\E}{\pa_r t_0}](t-t_0)|}.\label{1.21}
\end{eqnarray}
At the surface $t=t_0(r)$ we have a singularity, which corresponds
to the matter on this surface is passing through the stellar
center.
\begin{figure}[h]
\centering
\includegraphics[width=12cm]{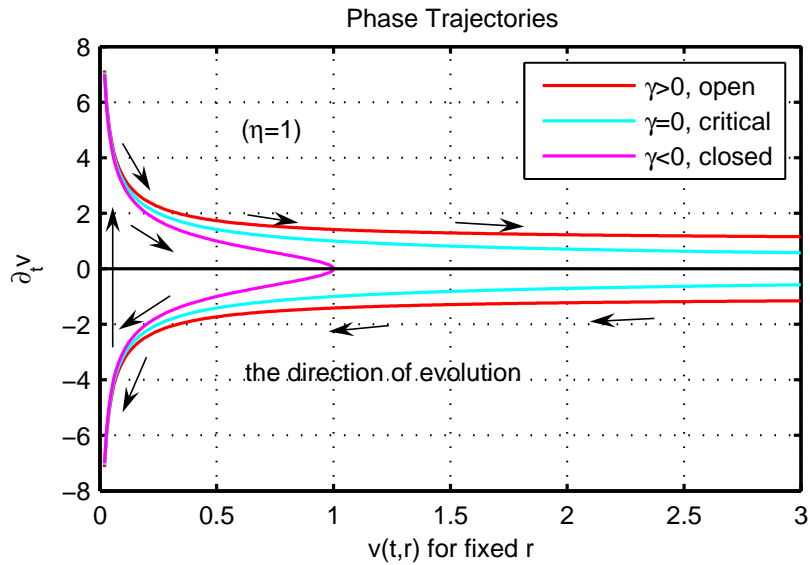}
\caption{There are three kind of trajectories for the free falling
dust. If $\ga<0$, which means the kinetic energy of the particles
is small, the orbits is oscillating. If $\ga\ge 0$, the orbits is
open} \label{fig}
\end{figure}
This space-time is bouncing. The phase trajectories of
(\ref{1.16}) are displayed in Fig(\ref{fig}). As shown below, if
$\ga < 0$, the space-time is oscillating, and if $\ga\ge 0$, the
space-time is bouncing.

In the case $\ga<0$, letting $\ga=-\frac \E {\eta^{2}}$, then we
get the solution
\begin{eqnarray}
\frac1 2 \eta^2\arccos\frac{\eta^2-2v}{\eta^2}- \sqrt{v(\eta^2-
v)}=\frac {\sqrt{\E}} \eta |t-t_0|. \label{1.17}
\end{eqnarray}
The space-time is oscillating. Solving $\pa_r v$ from (\ref{1.17})
and substituting it into (\ref{1.9}), we get
\begin{eqnarray}
{e^{\frac {\bar w} 2}}=\frac{ \sqrt{v(\eta^2-v)}}{2 \eta v
\sqrt{\E(\eta^2-\E)} }|[(\eta \pa_r\E -6\E\pa_r\eta)|t-t_0|
-2\sgn(t-t_0)\E\eta\pa_r t_0] +4\sqrt\E v^2\eta \pa_r\eta  |.
\end{eqnarray}

In the case $\ga>0$, letting $\ga=\frac {\E} {\eta^{2}}$, we get
the solution
\begin{eqnarray}
\eta^2 \ln\frac{\eta }{\sqrt{v}+\sqrt{\eta^2+ v}}+ \sqrt{v(\eta^2+
v)}=\frac {\sqrt{\E}} \eta |t-t_0|. \label{1.18}
\end{eqnarray}
The space-time is bouncing. Solving $\pa_r v$ from (\ref{1.18})
and substituting it into (\ref{1.9}), we get
\begin{eqnarray}
{e^{\frac {\bar w} 2}}=\frac{ \sqrt{v(\eta^2+v)}}{2 \eta v
\sqrt{\E(\eta^2+\E)} }|[(\eta \pa_r\E -6\E\pa_r\eta)|t-t_0|
-2\sgn(t-t_0)\E\eta\pa_r t_0] +4\sqrt\E v^2\eta \pa_r\eta  |.
\end{eqnarray}
When $t \to t_0$, the variables have the following asymptotic
behavior in all cases,
\begin{eqnarray}
v &=& e^{\frac w 2}\to \frac {1}{2}\sqrt[3]{18{\E}|t-t_0(r)|^2}. \\
e^{\frac {\bar w} 2} &\to& \frac{|\pa_r\E(t-t_0)-2\E\pa_r
t_0|}{\sqrt[3]{{12\E^2}|t-t_0|}\sqrt{1+\ga}} . \\
\rho &\to& \frac {4{\pa_r{\E}}\sqrt{1+\ga}}{3\kp
|[{\pa_r{\E}}(t-t_0)-2{\E}{\pa_r t_0}](t-t_0)|}.\label{1.30}
\end{eqnarray}

From the above calculation, we find the singularity at $t=t_0$
implies the free falling particles pass across the stellar center,
so it is a singularity caused by simplification of the matter
model, but not a real one. The dynamics for the pressure-free
fluid is ill-posed in mathematics, which results in the zero speed
of sound and `$\dl$-wave solutions'. So we can not take such
unrealistic model and solutions too serious, it just provides some
intuitions.

There are another kind singularity hides in the solutions. In the
critical case $\ga=0$, (\ref{1.20}) and (\ref{1.21}) become
singular at time
\begin{eqnarray}
t=t_0-\frac{2{\E}{\pa_r t_0}}{\pa_r{\E}}.\label{1.31}
\end{eqnarray}
If $t_0(r)$ is not a constant, then $t\ne t_0$, the singularity is
different from the one mentioned above.  $t_0(r)=const$ means a
moving mass shell. So whether this singularity is caused by the
globally invalidity of the comoving coordinate system or by the
ill-posed model is not clear. Strangely enough, the function $w$
or $v$ is normal at this time.

In the comoving coordinate system, the spatial coordinates are
defined by the trajectories of a set of particles, and the
temporal coordinate is defined by the proper time of the
particles\cite{wein,Oppen,Tolm}. In cosmology, since the space is
isotropic and homogeneous, this coordinate system works quite
well. However in the heavily curved space-time, the Gaussian
normal coordinate system is usually valid locally, and the global
simultaneous hypersurface can not be arbitrarily
defined\cite{gu8}. (\ref{1.14}) shows that, the comoving
coordinate system may be only valid locally. In what follows, we
examine the transformation between normal coordinate system and
the comoving one.

In general, the line element of an evolving space-time with
spherical symmetry is given by
\begin{eqnarray}
ds^2=b d t^{~2}+ 2f d t dr - a dr^2-r^2(d\th^2+\sin^2\th
d\vf^2),\label{2.1}
\end{eqnarray}
where $(a,b,f)$ are smooth functions of $(t,r)$. Making a
transformation for $t$ by solving a first order linear
differential equation, we can remove $f$ from the
metric\cite{wein}. This is equivalent to set $f=0$ in (\ref{2.1}).
Since for a normal star, the metric functions $(a,b,f)$ are at
least continuous functions with bounded first order
derivatives\cite{gu1}, then the solutions of the first order
linear differential for the coordinate transformations, such as
the one given in \cite{wein,gu2,gu}, have bounded second order
derivatives, so they provide globally valid coordinate
transformations. But the following analysis shows we usually can
not find the global comoving coordinate system in an arbitrary
regular space-time.

For any smooth metric functions $(a,b)$ with $\pa_r b\ne 0$,
assume the following transformation
\begin{eqnarray}
T = T(t,r),\qquad R= R(t,r)\label{2.2}
\end{eqnarray}
makes a comoving coordinate system with line element
\begin{eqnarray}
ds^2= d T^2-e^{\bar w}d R^2-e^w(d\th^2+\sin^2\th d\vf^2).
\label{1.1*}
\end{eqnarray}
Substituting (\ref{2.2}) into (\ref{1.1*}), and compare the
results with (\ref{2.1}) while $f=0$, we get the equations for
$(T,R)$ as follows
\begin{eqnarray}
((\pa_r T)^2+a)\pa_t R-\pa_r T \pa_t T\pa_r R =
0,\\
\pa_r  T \pa_t T \pa_t R - ((\pa_t T)^2-b)\pa_r R = 0. \label{2.5}
\end{eqnarray}
They are a linear equation system for parameters $(\pa_t R,\pa_r
R)$. The sufficient and necessary condition for nonzero solution
is the determinant of the coefficient matrix vanishes, then we
have
\begin{eqnarray}
a(\pa_t T)^2-b(\pa_r T)^2=ab . \label{2.6}
\end{eqnarray}
Let $u = -\pa_t R/\pa_r R$, substituting it into (\ref{2.5}) and
combining the equation with (\ref{2.6}), we get
\begin{eqnarray}
\pa_t T =\frac b {\sqrt{b-au^2}},\qquad \pa_r T =\frac {\pm au}
{\sqrt{b-au^2}}. \label{2.7}
\end{eqnarray}
In (\ref{2.7}), the signature $\pm$ can be removed by a
transformation $u\to-u$, so we only need to consider the positive
case. The integrable condition is given by
\begin{eqnarray}
\pa_t u   + u \pa_r u =\frac { \pa_t a }{2b }u^{3 }+ \l(\frac
{\pa_r b }{b   }-\frac {\pa_r a }{2a   }\r) u^{2}+ \l( \frac
{\pa_t b }{2 b }-{\frac {\pa_t a   }{a }}\r) u
   -{\frac {\pa_r b
}{2 a   }} . \label{2.8}
\end{eqnarray}
Each solution $u$ defines a comoving coordinate system. However,
this is a 1+1 dimensional nonlinear hydrodynamical equation, which
usually has only local smooth solution. The solution usually blows
up in a finite time even for the smooth functions $(a,b)$.

In summary, from the above calculation and analysis, we find that
the gravitational contraction is not so simple as it looks like.
The comoving coordinate system for the stellar system is only
compatible with the zero-pressure free-falling particles, so it is
not globally valid coordinate chart. The valid coordinate
transformation should be a 1-1 mapping with at least bounded
second order derivatives in its domain of definition\cite{gu}. The
free falling dust is just an idealized model, we can not treat it
too seriously. The space-time itself becoming singular lags behind
the matter source becoming singular, and then the conservative
property of the gravitational potential may resist the formation
of singularity\cite{gu}.

\section*{Acknowledgments}
The author is grateful to his supervisor Prof. Ta-Tsien Li and
Prof. Han-Ji Shang for their encouragement.

\end{document}